\documentclass[a4paper,11pt]{article}

\usepackage{graphics,color,array,dcolumn}
\usepackage{graphicx}
\usepackage{amssymb}
\usepackage{xspace}
\usepackage{color}

\usepackage{amsfonts}
\usepackage{latexsym}
\usepackage{amsmath}
\usepackage{vmargin}

   \def\be{\begin{equation}}
   \def\ee{\end{equation}}
   \def\ba{\begin{eqnarray}}
   \def\ea{\end{eqnarray}}

   \def\half{{1\over 2}}

   \newcommand{\bp}{{\bar{\psi}}}

   \newcommand{\fss}[1]{#1\!\!\!/}
   \newcommand{\cin}{i \fss{\partial}}

\setmarginsrb{30mm}{20mm}{30mm}{20mm}%
             {0mm}{10mm}{0mm}{10mm}
\title{Functional RG flow equation: regularization and  coarse-graining in phase space}
\author{G. P. Vacca\thanks{vacca@bo.infn.it} \ and 
        L. Zambelli\thanks{Luca.Zambelli@bo.infn.it}\\ 
\\
Dip. di Fisica, Universit\`a degli Studi di Bologna\\
INFN Sez. di Bologna\\
via Irnerio 46,  I-40126 Bologna,  Italy}
\date{}

\begin{document}
\maketitle

\begin{abstract}
Starting from the basic path integral in phase space 
we reconsider the functional approach to the RG flow of the one particle 
irreducible effective average action. 
On employing a balanced coarse-graining procedure for the canonical variables
we obtain a functional integral with a non trivial measure which leads to a
modified flow equation. We first address quantum mechanics for boson and
fermion degrees of freedom and we then extend the construction to quantum field theories. 
For this modified flow equation we discuss the reconstruction of the bare action 
and the implications on the computation of the vacuum energy density.
\end{abstract}

\section{Introduction}
Renormalization analysis in quantum (and statistical) field theory is one of
the main tools at our disposal to investigate non trivial theories. Both
perturbative and non perturbative techniques have been developed. Some
analytical results can be obtained starting from a functional
formulation, where the generator of the correlation functions is written in
terms of a functional (path) integral.

The Wilsonian idea of renormalization~\cite{Wilson}, which started from the analysis of
Kadanoff's blocking and scaling of spin systems (more generally
coarse-graining), can also be conveniently formulated analytically in a functional integral
formulation. The idea of a step by step integration
of the quantum fluctuations typically belonging to a momentum shell, followed by rescaling,
can be implemented in a smooth way~\cite{WH,Pol} and leads to a differential equation for 
 Wilson's effective action as a function of the scale parameter.
Another related approach is based on the study of the
renormalization group (RG) flow of the effective average action (EAA)~\cite{wetterich}.
Such an approach is a very powerful tool in the analysis of condensed matter
and quantum field theory (QFT) systems.

The starting point of all these formalisms is the functional quantization procedure 
based on the path integral in phase space with a Liouville measure corresponding to
the configuration variables ``$\!Q$" and their conjugate canonical momenta ``$\!P$".
The integrand depends on an action which is built from a Hamiltonian and thus
at this level the approach is not manifestly Lorentz covariant. 
For Hamiltonians quadratically dependent on the conjugate momenta, the
integration in the ``$\!P$" variables is trivially performed 
getting infinite factors contributing to the functional measure, 
which usually in flat spacetime is constant and thus can be neglected in the 
computation of correlation functions.
Thus one is left with a covariant formulation with functional integrals in
configuration space.  
At this point one usually implements a coarse-graining on the configuration 
variables ``$\!Q$" only, introducing a smooth cutoff in the Lagrangian action.

About this framework the following observations can be made.
First, usually the insertion of a smooth cutoff is performed in such a way to
modify the Lagrangian action, but the configuration space functional measure
is left unchanged. In order to understand that this 
is an unbalanced regularization of the divergences of the path integral 
it is sufficient to recall that the path integral of quantum mechanics is finite, 
as can be seen performing a suitable skeletonization, with singular contributions coming
from the configuration space functional measure, i.e. from the integration in 
the canonical momenta, canceled by others appearing in the integration of the 
Lagrangian action in configuration space.

Moreover, from the more fundamental point of view of the phase space path
integral, such a cutoff scheme is equivalent to a regularization of the
integral over configuration variables, but not over the conjugate momenta.
In fact the idea of coarse-graining as ``integrating out" 
quantum fluctuations in the UV region is usually implemented in a functional integral 
after the integration of the conjugate momenta is performed.
In this way even if any degree of freedom is associated to a pair of ``$\!P$" 
and ``$\!Q$" variables, the integration of the  ``$\!P$" modes inside one shell
is performed well before the integration of the ``$\!Q$" modes of the same shell.
Thus the whole procedure is unbalanced. 

In this work we therefore propose to implement a {\it balanced}
coarse-graining procedure in phase space, by introducing a cutoff operator
which affects both configuration variables and conjugate momenta.
We first analyze quantum mechanics of a system of bosonic degrees of freedom
and then consider also the fermionic case.
In both cases we argue that, once one has chosen the cutoff to be applied to the ``$\!Q$" variables,
out of all possible cutoff operators that could be applied to the ``$\!P$" variables
only one choice is natural and balanced, because it respects 
the correspondence between the action and the functional measure 
(either their Hamiltonian or Lagrangian versions) and does
not alter the pattern of cancellations of divergences taking place when integrating the pair
of conjugate variables.
It turns out that this choice corresponds to the simple regularization prescription
of introducing a scale and mode dependent operator in the symplectic structure of
the theory.
After such a modification of the functional integral in phase space,
we can then choose to integrate out the conjugate momenta (for a quadratic
dependence) and obtain a path integral in configuration space with a 
non trivial dependence of the measure on the cutoff operator.
According to this regularization, we investigate the RG flow equation for the
EAA, which is slightly modified with respect to the one obtained with the
usual procedure. This new flow equation, in contrast with the usual one, leads to the same results
obtained by other quantization methods.

Later we move to QFT, where the choice of the operator implementing the
coarse-graining procedure is guided by the requirement of Lorentz invariance. 
Nevertheless we briefly discuss also some consequences of Lorentz breaking choices.

Finally we illustrate in some cases the relation between the EAA satisfying our 
approach and the bare action defined as the one appearing inside the path integral, 
with or without a UV cutoff in the theory. 
In the UV region the approach of the EAA to the bare action is found 
to depend on the choice of the coarse-graining operator, and in particular 
on its singularity properties.

During our discussion we also analyze the flow of the constant term in the
potential of the EAA, which is related to the vacuum energy.
We anticipate that the Lorentz invariant coarse-graining leads naturally to 
an interesting fact: in presence of an UV cutoff $\Lambda$ the vacuum energy, 
which is computed integrating the flow from such an UV scale down to the IR,
is quadratically (and not quartically) divergent in $\Lambda$ for a free 
massive theory, whilst vanishes for a free massless theory. 
This property of the vacuum energy density was recently discussed~\cite{akhmedov,OS} 
on different grounds in a standard perturbative QFT framework,
by performing ad hoc subtractions justified by symmetry and reality conditions.

\section{Quantum mechanics}

\subsection{Bosonic degrees of freedom}
Let's consider a classical system with one bosonic degree of freedom
governed by the following Hamiltonian:
\be
 H(p,q) = \frac{1}{2}\, p^2 + V(q)
\ee
where $q$ and $p$ are canonically conjugate variables with 
Poisson bracket: $[q,p]=1$\ .
The quantization of such a system is performed via the following 
Euclidean phase-space path integral ($\hbar=1$): 
\be\label{hampathint1}
Z[J]=\int \left[ dp dq\right]\mu[p,q] e^{\int\! dt\, \left[i p(t)\partial_t
    q(t) - H\left( p(t),q(t) \right) + J(t) q(t)\right]}\ .
\ee
Here we explicited the presence of an unspecified functional measure $\mu[p,q]$.
In a skeletonized version of the path integral one usually considers $N$ time-slices  
and at each instant of time integrates over the corresponding phase space. 
Since the correct measure for each of these phase-space integrations is the 
Liouville measure, that is the square root of the determinant of the
symplectic form, in a time-slicing definition of the path integral the functional measure 
is the product of $N$ Liouville measures, which is clearly ill-defined in the
continuum limit $N\rightarrow\infty$.
Of course this is not the only possible source of infinities for $Z$: 
both the $p$ and the subsequent $q$ integration bring ill-defined factors. 
Since $Z$ could be UV-finite, as e.g. in the free particle case, we must
conclude that all these divergences can mutually cancel.
This leads to think that if one regularizes one of the contributions, 
also the others should be regularized in a consistent way.
This is not what is done in the manipulations of the 
functional integral leading to the many exact RG flow equations present in the literature.
In the following we are going to explain why this is so, restricting ourselves to the one
particle irreducible framework.
The translation of our reasoning to the flow of Wilson's effective action should be straightforward.

The usual modified configuration-space path integral lying behind the
Wetterich equation~\cite{wetterich} reads:
\be\label{wetterichZJ}
Z_k[J]=e^{W_k[J]}=\int[dq]\mu[q] e^{- \left(S[q] +\Delta S_k[q]\right) + \int\!dt\, J q }\, .
\ee
where $\mu[q]$ is a k-independent Lagrangian measure, for our bosonic system
$S[q]$ is the time integral of:
\be\label{bosonL}
L(q(t),\dot{q}(t)) = \frac{1}{2}\left(\partial_t q(t)\right)^2 + V(q(t))
\ee
and in ($\ref{wetterichZJ}$) one adds to it an infrared cutoff term 
$\Delta S_k(q)= \frac{1}{2}\int q(t) \, R_k(-\partial_t^2) q(t)$
to allow for the integration of ($-\partial_t^2$)-modes only above some 
scale $k^2$.
In other words, one introduces the following regulator:
\be\label{regularizer}
\Delta_k (-\partial_t^2) := \frac{R_k(-\partial_t^2)}{-\partial_t^2}
\ee
in the kinetic term of the action, by means of the substitution:
\be\label{regularization1}
\frac{1}{2}\int dt \, \left(\partial_t q(t)\right)^2  \to
\frac{1}{2}\int dt \, \partial_t q(t)\left( 1+ \Delta_k(-\partial_t^2) \right) 
\partial_t q(t)\, .
\ee
In this way one affects the divergences arising from the integration of the exponential
factor $e^{-S[q]}$, but does not modify the infinite determinant implicit 
in the Lagrangian functional measure.
Equivalently, the modified generating functional $(\ref{wetterichZJ})$ can be obtained by
the following phase-space path integral:
\be\label{wetterichhampathint}
Z_k[J]=\int \left[ dp dq\right]\mu[p,q] e^{\int\! dt\, 
\left[i \sqrt{1+\Delta_k(-\partial_t^2) }p(t)\partial_t q(t) 
- H\left( p(t),q(t) \right) + J(t) q(t)\right]}
\ee
by completing the square in the exponential and then integrating in the
momenta, thus getting an infinite factor changing the Liouville measure
$\mu[p,q]$ into the Lagrangian measure $\mu[q]$. 
Later in this section we will write such a factor as:
\be\label{ratioofmeasures}
\frac{\mu[q]}{\mu[p,q]} = \left({\rm Det}(-\partial_t^2)\right)^\half
\ee
because for the system under consideration we can express this ratio 
in terms of the functional integral for a free particle\footnote{
Recall~\cite{kleinert} that in a time-slicing definition of the path integral
for a Green function such that $T=\int dt=N\epsilon$
we have $\mu[p,q]=1^N$, $\mu[q]=(2\pi\epsilon)^{-N/2}$. For a free particle 
$\langle q,T|0,0\rangle=
\mu[q]\left({\rm Det}(-\partial_t^2)\right)^{-\half}e^{-q^2/2T}=(2\pi T)^{-\half}e^{-q^2/2T}$ 
(in units $m=\hbar=1$). 
On the rhs of eqn.~($\ref{ratioofmeasures}$) there is no factor $(2\pi
T)^{-\half}$ because we consider a measure $\left[dp\,dq\right]$ 
for a partition function.}.
Therefore we see that whilst Wetterich's prescription affects the definition
of the integral over $q(t)$ it does nothing for the integral over $p(t)$ nor
for the Hamiltonian functional measure.
The last form ($\ref{wetterichhampathint}$) of Wetterich's modified generating
functional is suggestive because 
it allows the interpretation of the coarse graining procedure as a modified 
Legendre transform, i.e. a $k$-dependent definition of the bare Lagrangian 
corresponding to a fixed bare Hamiltonian.

This also suggests us a way to implement the previously described principle of regularizing 
on the same footing both the exponential factor and the measure: if according
to Wetterich's prescription we modify the Legendre transform, we should also
consistently modify the symplectic structure because the two are strictly tied together.
In fact, recall that the Legendre transform term $p(t)\partial_t q(t)dt$ is
just the pull back of the Liouville 1-form $\lambda=p dq$ by means of the 
trajectory-parameterizing map $\left(p(t),q(t)\right)$ and that the symplectic 
form is $\sigma=d\lambda$.
Thus if we substitute $\lambda \rightarrow\lambda_k =\sqrt{1+\Delta_k}\ \lambda$
we should also substitute $\sigma \rightarrow\sigma_k =\sqrt{1+\Delta_k}\ \sigma$ and
correspondingly $\mu=\left({\rm Det}\ \sigma\right)^\half \rightarrow
\mu_k=\left({\rm Det}\ \sigma_k\right)^\half$.
In the following, to simplify the notation, we will take advantage of the fact
that for our system the Liouville measure is a constant, equal to one for
canonical coordinates, and we will write the previous functional measure as 
$\mu_k=\left({\rm Det}(1+\Delta_k)\right)^\half$.

To sum up, as a regularization prescription we propose the introduction of a
frequency-dependence in the symplectic structure, leaving unaltered the
Hamiltonian and the phase space manifold. 
In the present case this leads to the following generating functional:
\be\label{hampathint1}
Z_k[J]=\int \left[ dp dq\right]\left({\rm Det} (1+\Delta_k)\right)^\half 
e^{\int\! dt\, \left[i \sqrt{1+\Delta_k}\, p\partial_t q 
- H\left( p,q \right) + J q\right]}\ .
\ee
To see that this modification of the path integral affects all possible
sources of divergences, i.e. the $p$-integration, the $q$-integration and the measure,
it is sufficient to change the $p$-integration variable:
\be\label{changevar}
 P(t)= (1+\Delta_k(-\partial_t^2))^{1\over4}p(t)-
i(1+\Delta_k(-\partial_t^2))^{3\over4}\partial_t q(t) 
\ee
and get:
\ba\label{hampathint2}
Z_k[J]&=&\int \left[ dP dq\right]\left({\rm Det} (1+\Delta_k)\right)^{1\over4} 
e^{-\left( S[q]+\Delta S_k[q]\right)+\int\! dt\, \left[
-\half \left(1+\Delta_k\right)^{-\half} P^2 + Jq \right]} \nonumber \\
&=&\int \left[dq\right]\left({\rm Det}(-\partial_t^2) {\rm Det}
(1+\Delta_k)\right)^{1\over2} 
e^{-\left(S[q]+\Delta S_k[q]\right)+\int\! dt\, Jq}
\ea
where we evaluated the integral over $P(t)$ as in eqn.(\ref{ratioofmeasures})
 but this time with a $\Delta_k$ correction.
In terms of the usual EAA, defined by
\be\label{defgamma}
\Gamma_k[\bar{q}]= \mathop{\min}_{J}\left(\int\! dt\, J\bar{q}- W_k[J] \right)
- \Delta S_k[\bar{q}] \,,
\ee
eqn.($\ref{hampathint2}$) entails the following modified Wetterich equation for the RG flow: 
 \be\label{bosonERGE}
\dot{\Gamma}_k=\frac{1}{2} {\rm Tr} \left[\left(\Gamma_k^{(2)}+R_k\right)^{-1}
\dot{R}_k\right] -
\frac{1}{2} {\rm Tr} \left[ \left(-\partial_t^2 + R_k\right)^{-1}
\dot{R}_k\right]
\ee
where the dot stands for differentiation with respect to $\log k$ and must not
be confused with a time derivative, always denoted as $\partial_t$.
The derivation of this equation is identical to the original work by Wetterich; 
in brief, using eqs.($\ref{defgamma}$,$\ref{hampathint2}$,$\ref{regularizer}$)
and recalling that the connected two point function is the inverse of the Hessian matrix 
for $\left(\Gamma_k+\Delta S_k\right)$, we can write:
\ba
\dot{\Gamma}_k[\bar{q}] \!\!&=&\!\! -\dot{W}_k\left[{\frac{\delta
      \Gamma_k}{\delta \bar{q}}}\right]
-k\partial_k \Delta S_k\left[\bar{q}\right] \nonumber\\
\!\!&=&\!\! \half\int\!
dt\,\langle\left(q-\bar{q}\right)(t)\dot{R}_k(-\partial_t^2)
\left(q-\bar{q}\right)(t)\rangle_{\frac{\delta
    \Gamma_k}{\delta \bar{q}}} - k\partial_k \log{\left({\rm Det} 
(-\partial_t^2+R_k)\right)^{1\over2}} \nonumber\\
\!\!&=&\!\! \half\iint\! dt dt^\prime\, \left[\big(\Gamma_k^{(2)}+R_k\big)^{-1}\!\! - 
\big(-\partial_t^2+R_k\big)^{-1}\right]\!\!(t,t^\prime)\, 
\dot{R}_k(-\partial_t^2)\delta(t-t^\prime)\, .\nonumber
\ea

Notice that the naked differential operator in the additional subtraction term 
in eq.($\ref{bosonERGE}$) seemingly breaks the invariance under constant field rescalings.
This is not the case because the general form of this new term is a $(\log k)$-derivative of the
logarithm of the regularized functional measure. 
Under rescalings of the fields inside the path integral, the functional measure correctly
transforms and so does the subtraction term.
For example, since in the present case 
$\mu_k[q]=\left({\rm Det}(-\partial_t^2+ R_k)\right)^\half$,
if $q=\sqrt{Z_k}\, q^\prime$ then 
$\mu_k[q^\prime]=\left({\rm Det}\, Z_k(-\partial_t^2+ R_k)\right)^\half$
such that the subtraction term becomes
$-\frac{1}{2} {\rm Tr} \left[ \left(Z_k(-\partial_t^2 + R_k)\right)^{-1}k\partial_k{(Z_k R_k)}\right]$.

Using eqn.($\ref{bosonERGE}$) in the case of a single harmonic oscillator 
 and integrating the flow from $k=\infty$ down to $k=0$, one
obtains $\omega/2$ for the energy of the vacuum.
This result was already derived in~\cite{gies} starting from the usual Wetterich equation
and adding to it a subtraction term interpreted as corresponding to an UV
counter-term in the bare action,
required to guarantee that for zero potential and frequency the ground state
energy of the oscillator be zero.
Our point of view is different in that we would like to have a flow equation representation
of quantum mechanics where the quantities are finite at all scales and
counter-terms are unnecessary. 
We find that this can be achievied with our construction by  requiring that the 
$k\rightarrow\Lambda$ limit of $\Gamma_k$ be the classical action we are quantizing,
free of any $R_k$ dependence.
For a general discussion on this point we refer the reader to section $4$.
In the following we will briefly sketch how to derive this result in the present context.
  
Let us consider a truncation for the effective action with
$V_k(q)=E_k+\frac{1}{2} \omega^2 q^2$, so that
$\Gamma_k^{(2)}=(-\partial_t^2+\omega^2)\delta(t-t')$. We shall consider the
UV ``initial conditions''
such that $E_{k=\infty}=0$ and look for $E_{k=0}$. We shall later comment on this
choice when discussing the relation between the EAA and the bare action.
On employing the so called
optimized~\cite{litim}  cutoff function $R_k(z)=(k^2-z)\theta(k^2-z)$
and switching to a Fourier representation of the operators, one obtains
\be
\int \!dt\, \dot{V}_k=\frac{1}{2} \int \!dt \int \frac{dE}{2\pi}
\theta(k^2-E^2) 2 k^2 \left[ \frac{1}{k^2+V_k''}-\frac{1}{k^2}\right]
\ee
which, after removing the ``volume" factor ($\int dt$) on both sides of the equation, leads to
\be
\partial_k {E}_k=\frac{1}{\pi}\frac{-\omega^2}{k^2+\omega^2} 
\Longrightarrow E_{k=0}=\frac{\omega}{2}\, .
\ee
Let us finally interpret this result analyzing directly 
 the integro-differential equation which is satisfied by the EAA:
\be\label{intdiffeq}
e^{-\Gamma_k[\bar{q}]}\!=
\!\!\! \int [d q]  \, \mu_k
\exp{\left(\!\!-S[q]+\!\! \int (q\!-\!\bar{q})\frac{\delta
    \Gamma_k[\bar{q}]}{\delta \bar{q}}
  -\frac{1}{2}\! \int (q\!-\!\bar{q})\hat{R}_k(q\!-\!\bar{q})
\!\right)} \ .
\ee
Since for a free theory one has an EAA $\Gamma_k[\bar{q}]=S[\bar{q}]+E_k$,
one finds, using a compact notation where ``$\cdot$'' stands for an integration,
\ba\label{freeEAA}
e^{-\Gamma_k[\bar{q}]}\!\!\!&=&
\!\!\! \int [d q]  \, \mu_k
\exp{\!\!\left(\!-\frac{1}{2}(q\!-\!\bar{q})\!\cdot\!
  (-\partial_t^2+R_k+\omega^2)\!\cdot\!(q\!-\!\bar{q})-\frac{1}{2}\bar{q}\!\cdot\!
  (-\partial_t^2+\omega^2)\!\cdot\! \bar{q}
\!\right)} \nonumber\\
&=&\left(\frac{{\rm Det}\left(-\partial_t^2+R_k\right)}
{{\rm Det}\left(-\partial_t^2+R_k+\omega^2\right)}\right)^{1/2} 
\exp{\left(-\frac{1}{2}\bar{q}\cdot
  (-\partial_t^2+\omega^2)\cdot \bar{q}
\!\right)} \, .
\ea
One then notes that the first factor in the last line of eqn. \eqref{freeEAA}
becomes $1$ in the $k\to\infty$ limit while for $k\to 0$ gives the expected zero energy
contribution $e^{-\int \!dt \frac{\omega}{2}}$. We remark that in order to obtain these
values in the UV and IR limit of $k$ the cutoff operator $R_k$ should probably
satisfy some regularity conditions. These are fulfilled for the previously
mentioned optimized cutoff and for the so called Callan-Symanzik cutoff
($R_k=k^2$), but not for discontinuous cutoffs such as $R_k(z)=k^2\theta(k^2-z)$.  
\subsection{Fermionic degrees of freedom}
In this section we will study a free system
whose Lagrangian variables are $n$ real Grassmann-valued functions of time:
$\left\{\theta^i(t)\right\}_{i=1,...,n}$\ , evolving according to the following Lagrangian:
\be\label{spinorL}
L(\theta(t),\partial_t\theta(t))=\frac{1}{2}\theta^i(t)i\partial_t\theta^j(t)\delta_{ij}.
\ee
Just like in the previous section we consider as a starting point
the quantization of this theory by means of a Hamiltonian path integral.
In building a phase space out of (\ref{spinorL}) we find $n$ second class primary constraints:
\be\label{constraints}
\chi_\alpha(t) := \pi_\alpha(t)+\frac{i}{2}\delta_{\alpha j}\theta^j(t)=0
\ee
which cause the canonical Hamiltonian to vanish.
The relevant phase space is the surface $\cal{S}$ defined by ($\ref{constraints}$), 
a complete set of independent coordinates on it
is given by $\theta^i$ and the functional integral is to be taken over 
all paths $\theta^i(t)$ lying on this surface.
The appropriate measure for functional integration over $\cal{S}$ is again 
the square root of the superdeterminant of the symplectic form on $\cal{S}$.
In presence of second class constraints and assuming that the whole phase space
is endowed with a symplectic structure $\sigma$, 
we can define a nondegenerate symplectic form $\tilde{\sigma}$ on the reduced phase space,
simply by restricting $\sigma$ to $\cal{S}$. 
As the inverse of $\sigma$ is the Poisson bracket $[\, ,\, ]$, 
the inverse of $\tilde{\sigma}$ is the Dirac bracket $[\, ,\, ]_{\sim}$,
which in the reduced phase space coordinates $\theta^i$ has components:
$[\theta^i,\theta^j]_{\sim}=-i\delta^{ij}=[\chi_i,\chi_j]$ .
(Everything we write about constrained systems is explained for example in~\cite{HT}.)
Thus the functional integral over the reduced phase space reads:
\be\label{thetaint1}
 Z = \int [d\theta] \mu[\theta] e^{-\half\int\! dt\, \theta^i(t)\partial_t\theta^j(t)i\delta_{ij}}.
\ee
Here the Lagrangian ($\ref{spinorL}$) emerges from the
$\partial_t\theta^j\pi_j$ term in phase space
after having solved the second class constraints, 
or equivalently after having performed the following integration over momenta:
\be\label{intoverpi}
Z =\int [d\theta d\pi]\mu[\theta,\pi]
\left(\prod_\alpha \delta[\chi_\alpha]\right)  e^{-\int\! dt\, \partial_t\theta^j\pi_j}\ .
\ee

Following the same coarse graining scheme explained in the previous section we modify 
the symplectic structure of the reduced phase space replacing $\tilde{\sigma}$
with $\tilde{\sigma}_k$ = $(1 + \Delta_k) \tilde{\sigma}$,
where the definition of $\Delta_k$ is the same of ($\ref{regularizer}$) 
but for the replacement of $-\partial_t^2$ with $i\partial_t$.
To be more precise, in the fermionic case one usually chooses $R_k$ in such a way that
$|i\partial_t+R_k(i\partial_t)|^2$ is a regularized kinetic operator for a
bosonic degree of freedom, therefore the $k$-dependent operator we are
introducing in the symplectic structure is the same of the bosonic case.
Correspondingly the functional measure becomes: 
$\mu_k=\left( {\rm SDet}\, \tilde{\sigma}_k\right)^\half = \mu\, 
\left({\rm SDet}(1 + \Delta_k)\right)^{n\over 2}$.
Then the modified path integral reads:
\ba\label{realGrassmannZk}
 Z_k &=& \int [d\theta d\pi]\mu[\theta,\pi] \left(\prod_\alpha \delta[\chi_\alpha]\right) 
 \left({\rm SDet}(1+\Delta_k)\right)^{n\over 2} e^{-\int\! dt\,
   (1+\Delta_k)\partial_t\theta^j\pi_j}
\nonumber \\
&=& \int [d\theta] \mu[\theta] \left({\rm SDet}(1+\Delta_k)\right)^{n\over 2} 
 e^{-\half\int\! dt\, \theta^i(t)(1+\Delta_k)\partial_t\theta^j(t)i\delta_{ij}}.
\ea
Such a k-dependence can be translated in the following equation for the usual
effective 
average action:
\be\label{GrassmannERGE}
\dot{\Gamma}_k=\frac{1}{2} {\rm STr} \left[\left(\Gamma_k^{(2)}+R_k\right)^{-1}
\dot{R}_k\right] -
\frac{1}{2} {\rm STr} \left[\left(i\partial_t + R_k\right)^{-1}
\dot{R}_k\right]
\ee
where the traces as usual count also the number of Lagrangian variables.
Note that the $\half$ factors on the rhs are consistent with the traditional
Wetterich equation for Fermi fields, since in our case we are dealing with
\textit{real} Grassmann variables.

The generalization of the previous discussion to the case of n complex 
Grassmann variables $\{\eta^i\}_{i=1,...,n}$ and to interacting systems
 is straightforward.
As long as the kinetic term of the Lagrangian is of first order in 
time-derivatives and real, such as for example in: 
$L=\bar{\eta}^i i\partial_t\eta^j \delta_{ij} + V(\eta^i,\bar{\eta}^j)$,
 we find 2n second class primary constraints:
$\{\chi_\alpha,\bar{\chi}_\alpha\}_{\alpha=1, ..., n}$. $\chi$ relates 
the conjugate momentum of $\eta$ to $\bar{\eta}$, whilst
$\bar{\chi}$ relates the conjugate momentum of $\bar{\eta}$ to $\eta$. 
In order to find the correct functional measure
we can just compute the matrix of the Poisson brackets of these constraints.
Since:
\be
[\chi_\alpha,\chi_\beta]=[\bar{\chi}_\alpha,\bar{\chi}_\beta]=0\, , \,\,\,\,\,\,\,\,\,\,\,\,\,\,\,
 [\chi_\alpha,\bar{\chi}_\beta]=[\bar{\chi}_\alpha,\chi_\beta]
\ee
then  $\left|{\rm SDet}\, \tilde{\sigma}_{ij}\right|^\half = 
\left|{\rm SDet}\left([\eta^i,\bar{\eta}^j]_\sim^{-1}\right)\right|\, $
therefore, if we do not count the complex conjugate of a bracket 
as an independent bracket, the $\half$ exponent of the 
superdeterminants gets simplified in all previous formulas.
As a consequence, applying the same regularization scheme of 
eqn.~(\ref{realGrassmannZk}) we are led to
an equation for the effective average action which is identical
to eqn.~(\ref{GrassmannERGE}) but without the $\half$ factors on the rhs.

 Such an equation can be used to compute the vacuum energy of a fermionic oscillator in
quantum mechanics. Pick a complex Grassmann variable $\eta$,
and investigate the following truncation ($\omega>0$):
\be
\Gamma_k[\eta]= \int dt\, \left(\eta^\ast i\partial_t\eta + \omega\eta^\ast \eta + E_k \right)\, .
\ee
Proceeding along the same lines as for the bosonic oscillator, one finds that
the quantum energy of the vacuum is $E_0=-\omega/2$,
i.e. again what one would have computed by means of canonical quantization.

\section{Quantum field theory}
In this section we want to generalize the previous discussion to the case of field theory.
Let's start with the example of a scalar field theory with (Euclidean) Lagrangian density:
\be
{\cal L}= \frac{1}{2}(\partial_0\phi)^2 +\frac{1}{2}|\nabla \phi|^2 + V(\phi)
\ee
or equivalently, defining the momentum conjugate variable $\pi$ w.r.t. $\phi$, with
Hamiltonian density:
\be
{\cal H}= \frac{1}{2}\pi^2 +\frac{1}{2}|\nabla \phi|^2+V(\phi) \, .
\ee
As in section 2.1, our starting point is the quantization of such a system by 
the usual (Euclidean) Hamiltonian path integral 
\be\label{QFTZ1}
Z=\int \left[ d\pi d\phi\right] \mu[\pi,\phi]
e^{-\int d^d x\, \left(-i \pi \partial_0\phi + {\cal H}\right)}\ .
\ee
Again, in a skeletonized version of the path integral one considers $N$ time-slices
and at each instant of time integrates over a corresponding phase space; therefore
also in this case the functional measure is related to the Liouville measure. 
Hence let's just perform the same modification of the Liouville form we introduced before:
\ba\label{QFTregularization1}
\int \pi \partial_0 \phi &\to&
\int \pi \left( 1+ \Delta_k\right)^\half \partial_0 \phi \nonumber \\
\mu[\pi,\phi] &\to&
\mu_k[\pi,\phi]= \left( {\rm Det} \left( 1+ \Delta_k\right) \right)^\half \mu[\pi,\phi]
\ea
where we still do not specify which differential operator $\Delta_k$ depends on.
Exactly the same manipulations we performed in equations 
($\ref{changevar}$,$\ref{hampathint2}$) show that also in this case such a
prescription is sufficient to affect simultaneously the measure and
the following two quadratic forms in the action:
\ba\label{QFTregularization1bis}
\frac{1}{2} \int \partial_0 \phi \partial_0 \phi &\to&
\frac{1}{2}\int \partial_0 \phi  \left( 1+ \Delta_k\right) \partial_0 \phi \nonumber \\
\frac{1}{2} \int \Pi^2 &\to&
\frac{1}{2}\int \Pi \left( 1+ \Delta_k\right)^{-\half} \Pi \nonumber
\ea
where $\Pi$ is defined in analogy with eqn.($\ref{changevar}$).
But the remaining quadratic form $\partial_i\phi\partial^i\phi$ is left unaffected by
($\ref{QFTregularization1}$).
Therefore we must supplement ($\ref{QFTregularization1}$) with a second regularization:
\be\label{QFTregularization2}
\frac{1}{2} \int \partial_i \phi \partial^i \phi \to
\frac{1}{2}\int \partial_i \phi  \left( 1+\tilde{\Delta}_k\right)
\partial^i \phi
\ee
for some $\tilde{\Delta}_k$ a priori independent of $\Delta_k$.
In conclusion the final modified path integral for a generic theory of one scalar field reads:
\ba\label{ZkQFT}
Z_k&=& \int \left[ d\pi d\phi\right] \mu[\pi,\phi]\left({\rm Det}(1+\Delta_k)\right)^\half
e^{\int d^d x\, \left(i \pi \left(1+\Delta_k\right)^\half\partial_0\phi - 
{\cal H} \right) - \Delta H_k} \nonumber \\
&=&\int \left[d\phi\right] \mu[\phi]\left({\rm Det}(1+\Delta_k)\right)^\half
e^{-\left(S[\phi] + \Delta S_k [\phi]\right)}
\ea
where we denoted:
\ba\label{deltas}
\Delta H_k &=&\frac{1}{2}\int \partial_i \phi \tilde{\Delta}_k \partial^i \phi\nonumber\\
\Delta S_k &=&\frac{1}{2}\int \partial_0 \phi \Delta_k \partial^0 \phi + \Delta H_k \ .
\ea
Regarding the freedom to independently choose $\Delta_k$ and $\tilde{\Delta}_k$
we shall discuss in the next subsections several choices one can make.

There are also possible alternative approaches which are leading directly to covariant results.
One could be to adopt the so called covariant Hamiltonian formalism for
classical field theory, in which one introduces $d$ conjugate momenta to $\phi$, one for each
partial derivative. In this formulation a regularization of the consequent polysymplectic 
structure would automatically provide a $\tilde{\Delta}_k=\Delta_k$. Such an
approach deserves further analysis to define the appropriate functional measure which
should be adopted in the corresponding Hamiltonian path integral.

Another could be to slightly modify our regularization prescription, 
starting not from the phase space but from the configuration space  path
integral. In this framework the introduction of a covariant  $k$-dependent
operator in the Lagrangian must be accompanied by a  similar modification of
the corresponding Lagrangian measure, which is  the reciprocal of the square
root of the determinant of the advanced  Green function (see for
example~\cite{dewitt}).
We shall not discuss more on this here.

On the base of common sense we expect the case of fermionic fields 
to differ only by two aspects: the fields will be Grassmann-valued and there will
be constraints.
The first point only changes determinants in superdeterminants; the second one must be
dealt with as in section 2.2, integrating over the reduced phase space and, since a Dirac
fermion comprehends two complex Grassmann variables, using as a measure 
$\left({\rm SDet}([\psi,\bp]_\sim^{-1})\right)^2$.
Following these lines one gets the following modified path integral:
\be
Z_k =\int[d\bp d\psi]\mu[\bp,\psi] {\rm SDet}(1+\Delta_k)  e^{- \left(S[\psi] +\Delta S_k[\psi]\right)}
\ee
where:
\be\label{fermiondelta}
\Delta S_k =\int \bp \Delta_k i\gamma^0 \partial_0 \psi + 
\int \bp \tilde{\Delta}_k i \gamma^i \partial_i \psi\ .
\ee

Now it is time to comment on the definition of the operators $\Delta_k$ and $\tilde{\Delta}_k$.
Different choices of them can be made,
specifying their forms as functions of the modes and also the differential
operator whose modes they depend on.
Let us analyze few examples.

\subsection{Lorentz invariant cutoff}
The most natural choice is to take a Lorentz invariant cutoff. For a pure scalar theory
it is sufficient to investigate the modes of the Lorentz scalar operator 
$-\Box\equiv -\partial_\mu \partial^\mu$,
and in order to preserve Lorentz symmetry we are compelled to choose
$\tilde{\Delta}_k(-\Box)=\Delta_k(-\Box)$.
As already recalled in section 2.1 the usual notation is:
 $\Delta_k = \frac{ R_k(-\Box)}{-\Box}$ with the cutoff function $R_k(z)$ enjoying all the
properties required to suppress the functional integration for $z\ll k^2$.
Thus in this case eqn. ($\ref{deltas}$) reduces to the more traditional form:
\be
\Delta S_k[\phi]=\frac{1}{2}\int d^dx\, \phi R_k(-\Box ) \phi\, .
\ee
Plugging this expression into ($\ref{ZkQFT}$), defining the effective average action as usual,
and taking the $k$-derivative one gets:
\be\label{qftbosonERGE}
\dot{\Gamma}_k=\frac{1}{2} {\rm Tr} \left[\left(\Gamma_k^{(2)}+R_k\right)^{-1}
\dot{R}_k\right] -
\frac{1}{2} {\rm Tr} \left[\left(-\Box+R_k\right)^{-1}
\dot{R}_k\right]\, .
\ee
The requirement of Lorentz invariance also leads to a similar equation for Dirac Fermions:
\be\label{qftfermionERGE}
\dot{\Gamma}_k= {\rm STr} \left[\left(\Gamma_k^{(2)}+R_k\right)^{-1}
\dot{R}_k\right] -
 {\rm STr} \left[\left(\cin + R_k\right)^{-1}
\dot{R}_k\right]\,.
\ee

Let us analyze the flow equation for the scalar field case, using an optimized
cutoff function $R_k(z)=(k^2-z)\theta(k^2-z)$.
In the local potential approximation and
neglecting anomalous dimensions for $d=4$ one finds
\be
\dot{V}_k(\phi)=\frac{k^4}{2 (4\pi)^{2}}
\left[\frac{1}{1+V_k''(\phi)/k^2}-1\right]\, .
\ee
This is valid without approximation for a free massive field with
$V_k''(\phi)=m^2$. In this case the only non trivially running parameter is the
field independent term $v$ of the potential which we expect to contribute to the
``vacuum energy", while the dimensionful mass $m$ is constant along the flow.

One may consider the case in which $\Lambda$ is the scale at which the bare
action is defined. Then integrating the flow from $k=\Lambda$  down to $k=0$ one obtains
\be
v_{k=0}-v_{k=\Lambda}=\frac{1}{4 (4\pi)^{2}}
\left[ m^2 \Lambda^2 -m^4\log{\left(1+\frac{\Lambda^2}{m^2}\right)}\right]\,.
\ee
However this is not a quantitatively trustable estimate of the difference
between the quantum vacuum energy
and the bare one at the cutoff $\Lambda$, for reasons that we shall discuss in detail in section $4$.
Nevertheless it qualitatively agrees with the result we will compute in that section, because it 
correctly shows that the contribution to $v$ from the quantum
fluctuations, with the above Lorentz invariant prescription, is positive and
diverges only quadratically as $\Lambda \rightarrow \infty$. Moreover it vanishes in the
case of a massless field. This possibility was indeed
discussed~\cite{akhmedov,OS} on different grounds in a standard QFT framework.
We note that in our approach this fact is indeed related to Lorentz
invariance, but it is also the consequence of having treated carefully the
measure in the path integral which is the starting point of the quantization procedure.
Indeed if one employs the traditional Wetterich equation obtained by regulating the
bare action only, and not the measure, then
$v_{k=0}-v_{k=\Lambda}$ has an extra contribution
$-\Lambda^4/\left(8\, (4\pi)^{2}\right)$
which leads to a negative value for $\Lambda \gg m$,
is quartic divergent with the Lorentz invariant cutoff, and does
not vanish in the massless case.
As already discussed for the harmonic oscillator, this extra contribution
can be interpreted as due to a deformation of the bare action at the scale 
$k=\Lambda$ by the regulator $\Delta S_{\Lambda}$, and washed away by means of
a controlled subtraction in the flow equation, which is tantamount to adding 
counter-terms to $\Gamma_{k=\Lambda}$.
The fact that in the present approach these counter-terms are not needed
leads to the expectation that something has changed in the relationship
between the EAA and the bare action, as we will discuss in section $4$.

\subsection{Lorentz breaking pseudo cutoffs}
We consider here a special case where the cutoff is such that it organizes
the integration in the path integral according to the modes associated to
the operator $-\partial_t^2$.
More precisely we employ the same $\Delta_k( -\partial_t^2)$ of
eqn.~(\ref{regularizer}) and we take a vanishing $\tilde{\Delta}_k$.
Therefore the quadratic form in the spatial derivatives of the field is not affected at all
and any integration in the $(d-1)$-dimensional space is divergent and should
be regularized by other means. The flow equation now reads
\be
\dot{\Gamma}_k=\frac{1}{2} {\rm Tr} \left[\left(\Gamma_k^{(2)}+R_k\right)^{-1}
\dot{R}_k\right] -
\frac{1}{2} {\rm Tr} \left[\left(-\partial_t^2+R_k\right)^{-1}
\dot{R}_k\right]\, .
\ee
In the Fourier representation and employing the same optimized cutoff
function $R_k(z)$ as before, one can write
\be
\dot{V}_k(\phi)=\frac{k^3}{\pi}\int \frac{d^{d-1}\bar{p}}{(2\pi)^{d-1}}
\left[\frac{1}{k^2+|\bar{p}|^2+V_k''(\phi)}-\frac{1}{k^2}\right]\, .
\ee
Again for a free massive field, assuming an implicit regularization for the
$\bar{p}$ integration one can perform the integration of the flow and,
defining $\omega_{\bar{p}}=\sqrt{|\bar{p}|^2+m^2}$, obtains
\be
v_{k=0}-v_{k=\Lambda}=\!\!\!\int_{reg} \!\frac{d^{d-1}\bar{p}}{(2\pi)^{d-1}}
\frac{\omega_{\bar{p}}}{\pi}
\arctan{\frac{\Lambda}{\omega_{\bar{p}}}}
\underset{\Lambda\to \infty}{\longrightarrow}
\int_{reg} \!\frac{d^{d-1}\bar{p}}{(2\pi)^{d-1}}
\frac{\omega_{\bar{p}}}{2}
\ee
which is the usual integrated vacuum energy of all the vacuum fluctuations,
which is not a Lorentz scalar. Note that the original Wetterich equation for such a cutoff
leads to an extra divergent negative contribution:
$-\frac{\Lambda}{\pi}\int \!\!\frac{d^{d-1}\bar{p}}{(2\pi)^{d-1}}$.

Other kinds of Lorentz-breaking coarse-graining procedure can be implemented
using more complicated cutoff operators. As a simple non trivial example consider  
$R_k(-\partial_t^2,-\Box)=(k^2-(-\partial_t^2))\theta[k^2-(-\Box)]$ for 
$\Delta_k=1+R_k(-\partial_t^2,-\Box)/(-\partial_t^2)\, $, while $\tilde{\Delta}_k=0$ .
This cutoff organizes the integration of the modes according to the
eigenvalues of the Laplacian, which is
a Lorentz invariant operator, but modifies only the time derivative part in
the action (and the conjugate momenta) leaving the spatial derivative term
untouched.

Let us note that sometimes it may be useful from a phenomenological point of
view to allow for a Lorentz-breaking coarse-graining procedure, depending on
which observable one may be interested in and on the experimental setup.

\section{Relation between the EAA and the bare action}
An interesting point to be discussed, already addressed in~\cite{RM} for the standard
Wetterich equation, is the relation between the EAA
satisfying the flow equation and the bare action of the
theory, appearing inside the functional integral. 
Traditionally physicists have been interested in investigating this relationship 
only in one direction, i.e. moving from the choice of a ``classical" bare
action to the computation of the corresponding ``quantum" effective action,
i.e. in flowing the RG towards the IR.
Why should the other direction being interesting?

One answer could be that if we turn the previous point of view upside down,
looking for the action to be plugged inside a path integral in order to get a previously chosen 
quantum effective action, we are just looking for a Wilson effective action, 
whose scale of reference depends on the scale we use to regularize the path integral.
Thus, in the limit in which this regularization is removed, the bare action we are looking for
becomes the UV limit of Wilson's effective action.
This is why the RG flow of the EAA towards the UV has been used in many recent
studies devoted to investigate the possible UV completion of several QFTs,
in the sense of Weinberg's asymptotic safety~\cite{weinberg}.
In particular, many efforts were directed to the case of gravity~\cite{reuterAS}.

Another possible answer, as already stressed in~\cite{RM}, is that 
 finding which classical system, once quantized, leads to the
theory under investigation, is important to establish relationships with other theories that
should describe the same system but that follow from different quantization schemes.
For example, in asymptotic safety scenarios for gravity, knowledge of the bare
action might lead to a better understanding of possible relations between the
QFT defined by renormalizable trajectories in the theory space of Einstein
gravity and other approaches to quantum gravity.

Thus we face the problem of computing the bare action from the EAA
just by means of the flow equation and its solutions, without resorting to the path
integral formulation of QFT.
In this section for sake of simplicity we will address this problem only for a scalar theory,
restricting ourselves to a Lorentz invariant cutoff $R_k(-\Box)$ (or
$R_k(-\partial_t^2)$ in the QM case).
We shall see that in the present case of a modified Wetterich equation we can
push the analysis of the relation between the EAA and the bare action in a
slightly different direction from the standard approach.
We organize our discussion distinguishing between two qualitatively different cases, 
the one in which a sharp UV cutoff is introduced, and the one without it.
No need to recall that the latter is allowed because of the ERGE being 
free of UV divergences.

\subsection{In presence of a UV cutoff}

So, let's assume that our theory has a Lagrangian bare action $S^\Lambda$ defined in the
presence of a configuration space measure $\mu^\Lambda$, both dependent on a 
UV cutoff $\Lambda$. 
Of course one is free to redefine the bare action adding the
$-\log{\mu^\Lambda}$ term to $S^\Lambda$ and removing $\mu^\Lambda$ from the measure.
The dependence on the UV cutoff $\Lambda$ can also be seen as
reflecting the fact that our bare action might follow from a coarse-graining procedure
started with some other bare action defined on a larger space
with a larger cutoff, in the Wilsonian sense. Therefore for different values
of $\Lambda$ one has a set of different
Wilsonian actions $S^{\Lambda}$ all referring to the same physical system. 
The removal of the UV cutoff is
associated to the limiting procedure $\Lambda\to \infty$, and is possible only
for fundamental, in contrast to effective, theories.
Starting from a path integral of the kind
\be
\int [d \chi]^\Lambda \, \mu^\Lambda e^{-S^\Lambda[\chi]}
\ee
for fundamental theories one should obtain finite meaningful matrix elements 
in the limit $\Lambda\to\infty$.

In presence of both the IR cutoff $k$ and the UV cutoff $\Lambda$, 
the definition of the EAA $\Gamma_k^\Lambda$ is formally the same of
the case without any $\Lambda$. As before, under the requirement that 
$R_k$ vanishes in the limit $k\to 0$, $\Gamma_k^\Lambda$ approaches the
standard effective action $\Gamma^\Lambda$ in the same limit.
A less simple problem is what happens to $\Gamma_k^\Lambda$ as $k$ grows.
Regardless of the presence of any UV cutoff, since as $k$ becomes bigger and bigger 
less and less modes of the fields are being integrated,
the most reasonable requirement seems to be that when the functional integration is completely
suppressed, the EAA approaches the bare action.
The best way to understand how this can happen is to look at the
integro-differential equation satisfied by the EAA, which in our case depends on a
regulated configuration space measure $\mu_k^\Lambda$:
\be\label{integrodiff}
e^{-\Gamma_k^\Lambda[\phi]}\!=
\!\!\! \int [d \chi]^\Lambda  \, \mu_k^\Lambda
\exp{\left(\!\!-S^\Lambda[\chi]+\!\! \int^\Lambda (\chi\!-\!\phi)\frac{\delta
    \Gamma_k^\Lambda[\phi]}{\delta\phi}
  -\frac{1}{2}\! \int^\Lambda (\chi\!-\!\phi)\hat{R}_k(\chi\!-\!\phi)
\!\right)} \ .
\ee
If, as $k$ grows and approaches some limiting value, the $k$-dependent part 
of the integrand on the rhs of eq.(\ref{integrodiff})
converges to a representation of a functional delta
\be
\mu_k^\Lambda\exp{\left( \int^\Lambda (\chi\!-\!\phi)\frac{\delta
    \Gamma_k^\Lambda[\phi]}{\delta\phi}
-\frac{1}{2}\int^\Lambda (\chi\!-\!\phi)\hat{R}_k(\chi\!-\!\phi)\right)}
\longrightarrow \delta^\Lambda[\chi-\phi]
\label{delta_rep}
\ee
then at this limiting value of $k$ the functional integral is completely
suppressed and the EAA equals the bare action.
To ensure that the $k$-dependent terms define a rising delta functional
we need to make assumptions both on the UV asymptotics of the EAA and on the 
properties of the regulator $R_k$. In fact the lhs of eqn.~(\ref{delta_rep})
 can be rewritten, in condensed notation and by completing the quadratic form, as
\be\label{completesquare}
\mu_k^\Lambda \exp{
\left\{\frac{1}{2}\frac{\delta \Gamma_k^\Lambda[\phi]}{\delta\phi} \cdot \hat{R}_k^{-1}\cdot
\frac{\delta \Gamma_k^\Lambda[\phi]}{\delta\phi}
\! -\half 
\left(\! \chi\!-\!\phi \!-\! \frac{\delta \Gamma_k^\Lambda[\phi]}{\delta\phi}
\cdot \hat{R}_k^{-1}\!\right) 
\cdot \hat{R}_k\cdot
 \left(\! \chi\!-\!\phi \!-\! \hat{R}_k^{-1}\cdot \frac{\delta 
\Gamma_k^\Lambda[\phi]}{\delta\phi}\right)
\right\} }\ .
\ee
>From this last equation we see that if the first term in the exponent of 
eqn. (\ref{completesquare}) and the shift term 
$\hat{R}_k^{-1}\cdot \frac{\delta \Gamma_k^\Lambda[\phi]}{\delta\phi}$ both
vanish when $k$ reaches its limiting UV value, and if the remaining functional
\be\label{gaussian}
\mu_k^\Lambda \exp{
\left\{ 
-\half 
\left(\! \chi\!-\!\phi\!\right) 
\cdot \hat{R}_k\cdot
 \left(\! \chi\!-\!\phi\!\right)
\right\} }
\ee
behaves as a normalized Gaussian functional with vanishing variance in the same limit, 
this is enough to recover a delta functional. These conditions
can on their turn be satisfied by assuming that physically allowed EAA are
bounded in $k$ and that the dimensionless version of $R_k$ diverges as $k$
reaches its limiting UV value.

Of course the details about which limiting UV value of $k$ one could approach 
and of which properties the regulator $R_k$ must enjoy
in order to completely suppress the integration and furnish the required rising delta
strongly depend on the presence or absence of a UV cutoff.
In the following we shall comment on all these details and also on the
assumption that the EAA be bounded in $k$,
but before starting analyzing all possible scenarios let us stress that
in order to have a chance to build a rising delta functional a crucial role is played
by the regularized functional measure (corresponding to a regularized
Liouville measure in phase space), without which we would lack the Gaussian 
normalization factor in ($\ref{gaussian}$).
Later on we will explicitly work out in a simple specific case the proof
that our regularization of the functional measure is exactly what is needed to 
normalize the Gaussian rising delta.
Finally it could be useful to recall that in studying the UV asymptotics, 
and in other computations too, one should choose a unit of mass and work with 
dimensionless quantities, that is, one should perform a general rescaling 
with respect to some $M$.
In presence of a UV cutoff $\Lambda$, $M=\Lambda$ is a possible choice related to the
domain of definition of our theory. If the UV cutoff $\Lambda$ is absent $M=k$ 
is also a natural choice. 

So far only general arguments, so let us start getting more specific.
Let us first deal with the case in which the presence of a UV cutoff $\Lambda$ 
is explicitly assumed. In this case there seem to exist two main choices
for the limiting value that $k$ must approach to suppress the integration: 

a) $k\to\Lambda$

b) $k\to\infty$
\noindent 

The former corresponds to the interpretation of $k$ as the scale of an IR
cutoff, that must therefore be smaller or equal to the UV cutoff. 
The latter is allowed since the $k$-dependent operator is a smooth IR
regulator and not a sharp cutoff, thus one might prefer to think about 
$k$ just as an external parameter that could take any value, 
in which case (b) corresponds to having moved to infinity the
arbitrary value of $k$ at which the regulator is expected to kill the integration.
The next question is which properties must $R_k$ enjoy in cases (a) and (b) 
in order to realize the two following scenarios:

a) $\displaystyle\lim_{k\to\Lambda} \Gamma_k^\Lambda[\phi]=S^\Lambda[\phi] $

b) $\displaystyle\lim_{k\to\infty} \Gamma_k^\Lambda[\phi]=S^\Lambda[\phi] $

\noindent 

Since case (b) is more similar to the case most frequently addressed in the 
literature, i.e. the one in which no UV cutoff is assumed, we prefer to start 
with this case. Here we want the dimensionless regulator $R_k$ to
diverge as $k\to\infty$ and, because in this case the most natural choice of 
the unit is $M=k$ for $k\leq\Lambda$ and $M=\Lambda$ for $k>\Lambda$, this singular behavior
is enjoyed by all regulators usually present in the literature, of the form 
\be\label{oldregulators}
\langle x| \hat{R}_k |y \rangle=k^2 f(-\Box_x/k^2)\delta(x-y)\;\;\,,\;\;f(0)>0
\ee
(the optimized cutoff corresponds to $f(z)=(1-z)\theta(1-z)$).
For such regulators and under the assumption of UV boundedness of $\Gamma_k^\Lambda$
one can easily check that for $k>\Lambda$,
rescaling all dimensionful quantities w.r.t. $M=\Lambda$,
 the first term in the exponent of eqn. (\ref{completesquare})
and the shift term $\hat{R}_k^{-1}\cdot \frac{\delta \Gamma_k^\Lambda[\phi]}{\delta\phi}$ both
vanish in the limit $k\to\infty$. 
Also, by the same token, the remaining exponential has a vanishing variance.
The last ingredient missing for a rising delta is the correct normalization, 
therefore it is time to give explicit arguments showing that this is provided 
by the regularized measure. To this end let us analyze briefly the QM case in the
skeletonized version of the path integral (see for example~\cite{kleinert})
 with a time slicing such that $T=N\epsilon$. 
We will not work out the exact discretized version of any cutoff differential 
operator, but we will just consider its asymptotic IR and UV behaviors.
In the $k\to 0$ limit, corresponding to the standard
un-coarse-grained path integral, $R_k$ and $\Delta_k$
disappear and for a standard unit Liouville measure the momenta integration
leads to the usual configuration space measure $\mu[q]={\cal N}=(2\pi \epsilon)^{-N/2}$,
as already discussed in footnote $1$.
In the case of $k \to \infty$, $-\partial_t^2/M^2$ is negligible with respect to $R_k/M^2$
for all modes, because of the cutoff $\Lambda$.
Also, having in mind a path integral toy-modeling a vacuum persistence amplitude,
we consider the case in which we have to perform $N$ integrals over phase space $(p,q)$,
so that the appropriate power of the regularization of the Liouville measure 
is also $N$. Therefore:
\be
\mu_k[q]\mathop{\sim}_{k\to\infty}{\cal N} 
\left(\frac{{\rm Det}[R_k/M^2]}{{\rm Det}[-\partial_t^2/M^2]}\right)^{1/2}\!\sim
(2\pi \epsilon M)^{-N/2} \frac{\left(\frac{R_k}{M^2}\right)^{N/2} }{
  (\epsilon M)^{-N}}= \left({\epsilon R_k\over 2\pi M}\right)^{N/2} \ .
\ee
Here the ${\rm Det}[-\partial_t^2]$ has been evaluated in terms of the finite
difference operators $\nabla$ and $\bar{\nabla}$ as in~\cite{kleinert}.
This measure and the satisfied requirement that the $k$-dependent exponent 
on the rhs of eqn.~(\ref{integrodiff}) approaches a Gaussian in the
$k\to\infty$ limit, leads to the following discretized
version of the l.h.s. of eqn.~\eqref{delta_rep}, or equivalently of ($\ref{gaussian}$)
\be
\left({\epsilon R_k\over 2\pi M} \right)^{N/2} e^{-\frac{1}{2M}\epsilon \sum_n R_k (q_n-\bar{q}_n)^2}
\underset{k \to \infty} \longrightarrow \prod_n
\delta(q_n-\bar{q}_n)
\label{delta_discrete}
\ee
showing the correct normalization.
Recall that in the last expression the rescaling has converted 
$q_n$ and $\bar{q}_n$ into dimensionless quantities and that in this
discretized approach $\epsilon$ is related to the inverse of the UV cutoff.  

In conclusion, for the class of operators ($\ref{oldregulators}$) and 
for any fixed bare action $S^\Lambda[\phi]$ at a given
scale $\Lambda$, one has 
$\displaystyle\lim_{k\to\infty} \Gamma_k^\Lambda[\phi]=S^\Lambda[\phi]\ $.
Thus, in order for the bare action in the limit $\Lambda\to\infty$ to be 
the initial condition at $k\to\infty$ for the RG flow of the EEA, 
one must deal with the following order in the limits: 
$\displaystyle\lim_{\Lambda \to \infty}\lim_{k \to \infty}
\Gamma_k^\Lambda[\phi]$.
The only hypothesis we still have to comment on, is the one regarding
 the boundedness in $k$ of $\Gamma_k^\Lambda$. With such an aim in mind
let us analyze the flow of the (modified) exact RG equation (ERGE) for
truncations like the local potential approximation. The
computation of the trace in Fourier space requires to integrate in $p$, 
over the domain $|p|<\Lambda$, a function $g(p,\Lambda,k,\phi)$ depending on all
the scales.  Rescaling everything w.r.t. $\Lambda$ ($z=p^2/\Lambda^2$ and 
$\tilde{\phi}=\phi/\Lambda$),
and adopting again an optimized cutoff as an example, on has for the ERGE:
\be
k\partial_k \,v_{k/\Lambda}(\tilde\phi)
=\frac{1}{(4\pi)^{\frac{d}{2}}\Gamma\left(\frac{d}{2}\right)}
\int^{min\{\frac{k^2}{\Lambda^2},1\}}_0 \!\!dz \,z^{\frac{d}{2}-1} 
\frac{-v_{k/\Lambda}''(\tilde\phi)} 
{\frac{k^2}{\Lambda^2}+v_{k/\Lambda}''(\tilde\phi)}
\ee
whose r.h.s. for a generic potential is expected to vanish when $\Lambda$ 
is fixed and $k\to\infty$. For example for a free massive theory
$v_{k/\Lambda}''(\tilde\phi)=m^2/\Lambda^2$, which is actually $k$-independent.
Therefore in such a case the EAA really 
approaches the $\Lambda$ dependent bare action.
Integrating the flow for the massive free theory from $k=\infty$ (instead of from
$k=\Lambda$) to $k=0$ one finds for the dimensionful energy density:
\ba\label{vacuumenergy}
\!\!\!&d=1&\!\!\!:  \,\,\,
 \frac{m \arctan\left(\frac{\Lambda }{m}\right)}{\pi }+
\frac{\Lambda  \log \left(1\!+\!\frac{m^2}{\Lambda ^2}\right)}{2 \pi }
\underset{\Lambda \to \infty}  \longrightarrow \frac{m}{2}\nonumber\\
\!\!\!&d=2&\!\!\!: \,\,\,
\frac{m^2 \log \left(\frac{\Lambda ^2}{m^2}\!+\!1\right)}{8 \pi }+
\frac{\Lambda ^2 \log \left(1\!+\!\frac{m^2}{\Lambda ^2}\right)}{8 \pi }
\underset{\Lambda \to \infty}  \longrightarrow
\frac{m^2 \log \left(\frac{\Lambda }{m}\right)}{4 \pi }+\frac{m^2}{8 \pi
}\\
\!\!\!&d=4&\!\!\!: \,\,\, 
\frac{m^2 \Lambda ^2-m^4 \log \left(\frac{\Lambda ^2}{m^2}\!+\!1\right)}{64 \pi
  ^2}+
\frac{\Lambda ^4 \log \left(1\!+\!\frac{m^2}{\Lambda ^2}\right)}{64 \pi ^2}
\underset{\Lambda \to \infty}  \longrightarrow
\frac{m^2 \Lambda ^2}{32 \pi ^2}-\frac{m^4 \log \left(\frac{\Lambda
  }{m}\right)}{32 \pi ^2}- \frac{m^4}{128 \pi^2}\, .\nonumber
\ea
>From the last expression on can see that in a four dimensional spacetime the term in the
Wilsonian action associated to the vacuum energy density induced by free
massive bosonic fields grows quadratically with the UV cutoff $\Lambda$ and is
zero for the massless case.
A similar behavior (but opposite sign) is shown by fermion fields.
We note that for a generic dimension $d$ one can write an analytic form as the
sum of two contributions:
one comes from the flow in the region $\Lambda<k<\infty$ and
is given by $\frac{\Lambda^d}{d \,\Gamma(d/2)
  (4\pi)^{d/2}}\log\left(1\!+\!\frac{m^2}{\Lambda ^2}\right)$ while the other is
obtained from the region $0<k<\Lambda$ and can in general be written in terms of 
a regularized hypergeometric function. More general truncations with higher derivative
terms or more complicated operators have to be studied to understand if 
such a behavior can be spoiled.

Let us now turn to case (a), again in presence of a finite UV cutoff. 
Here we want the dimensionless regulator $R_k$ to diverge as $k\to\Lambda$ from below.
Since the flow always stays in the region $k<\Lambda$ the choice of unit $M=k$ is allowed
and is in fact to be preferred to $M=\Lambda$ because rescalings with respect
to the running cutoff correspond to the Wilsonian procedure of iterated shell 
integration and subsequent rescaling. 
The cutoff function in general may be written as 
$R_k=k^2 f(z=\frac{p^2}{k^2},x=\frac{k^2}{\Lambda^2})$.  If $f(z,x)$ is {\it finite} at $x=1$ one 
does not obtain a Gaussian representation of the delta and
also both the first term and the shifts to $\chi - \phi$ in the exponent of
eqn.~(\ref{completesquare}) do not vanish. In this case there is no simple relation between
 $\Gamma_{k\to\Lambda}$ and $S$.
On the other hand if we ask that $f(z,x)\to \infty$ as $x\to 1$, and still we assume the
boundedness of the EAA,
then for any mode we recover a functional
delta representation as in eqn.~(\ref{delta_rep}). In such a case one may write
$\lim_{k\to\Lambda} \Gamma^\Lambda_k[\phi]=S^\Lambda[\phi] \, .$

One can check that for a free theory, independently on the choice of the
functional form providing the singularity in the cutoff function one obtains
the same results already illustrated in eq. \eqref{vacuumenergy}. This is a
confirmation that one has realized  a representation of the functional delta
inside the path integral. In particular we have considered the explicit
cutoff $R_k(p^2)=f(\frac{k^2}{\Lambda^2})(k^2\!-\!p^2)\theta(k^2\!-\!p^2)$. 
In order to perform the analytic computation
we used $f(x)=(1\!-\!x)^{-\alpha}$ for positive $\alpha$ and 
numerically checked that other suitable choices of $f$ lead to the same result. 
In this case therefore it should be possible to find a change of variable in
order to make this property manifest for any $f$ with the right singular behavior.  

We stress again that the difference between the scenarios realized in cases
(a) and (b) just depends on our freedom to choose cutoff operators with a
different singular behavior. The EAA depends on them at all scales but in the
UV and IR limit, provided such cutoffs enjoy regularity properties allowing to
bring the $k\to0$  limit inside the path integral, and that their
singular behavior in the UV is what is needed to recover a functional delta.

\subsection{In absence of any UV cutoff}

Finally we want to study the case in which no UV cutoff is present.
As already said this last case is the one which is usually considered in the EAA approach, 
with $\Gamma_k$ assumed to be bounded in $k$ because of renormalizability.
In this framework the most reasonable request is that
$$\displaystyle\lim_{k\to\infty} \Gamma_k[\phi]=S[\phi] $$
and one should look for cutoff operators leading to a representation of
the delta for $k\to\infty$. 
Since the only natural choice of unit in this case is $M=k$, we see that
the traditional regulators like ($\ref{oldregulators}$), after rescalings
with respect to $k$, do not diverge as $k\to\infty$ unless $f(0)$ is infinite.

Thus, an example of a suitable regulator could be
$R_k(p^2)=g(\frac{p^2}{k^2})(k^2-p^2)\theta(k^2-p^2)$ with $g(z)\to\infty$ for
$z\to 0$.
Let us remark that with this kind of choice the singularity $k \to \infty$ is 
the same as the one for $p\to0$ since $k$ is the only available scale in the cutoff. 
This means that the zero mode is treated differently inside the path integral.
The simplest computation one can imagine is to check that this prescription
leads to the expected vacuum energy in $d=1$ for a free theory with frequency
$m$ (which we know to be finite starting from a bare action with no
vacuum energy at $k=\infty$). Choosing $g(z)=z^{-1}$ one can compute analytically
the beta function from the trace (its expression is more involved than that for
the simpler optimized cutoff) and numerically check that the flow leads to
$E_{k=0}=\frac{m}{2}$. A full numerical analysis on a family of cutoff
defined, for example, by $g(z)=z^{-\alpha}$ for $\alpha>0$ gives correctly the
same result.

Another suitable family of cutoff functions, far easier to deal with, is 
$R_k(z k^2)=k^2 (z^{-\alpha}-z)\theta(1-z)$ for $\alpha>0$. These regulators simply
replace $(p^2/k^2)$ with $\left(p^2/k^2\right)^{-\alpha}$, whenever $p^2<k^2$.
The choice $\alpha=d/2$ allows a straightforward computation of the traces
thanks to the advantageous change of variable $z^\prime=z^{d/2}$. 
As an example, for a scalar field in the local potential approximation one finds:
\be
\dot{V}_k(\phi)=\frac{k^d}{ (4\pi)^{{d\over2}} } \frac{\left({d\over2}+1\right)}{\Gamma\!\left({d\over2}+1\right)}
\left[\frac{k^2}{V_k''(\phi)}\log\left(1+\frac{V_k''(\phi)}{k^2}\right) -1 \right]\, .
\ee
Again for $d=1$ we see the independence of the result $E_{k=0}=\frac{m}{2}$ from the choice of the cutoff function
with the required singularity structure.  
Clearly without a regularizing UV cutoff the integrated vacuum energy for $d>1$ is a
divergent quantity so one should only deal with the expressions for the beta functions. 
Again further investigations are needed regarding more general truncations of
the EAA. 
Finally let us remind to the reader that for the special case of a free theory we have shown in the end of
section $2.1$ (for the QM oscillator) that it is sufficient to employ even a non
singular cutoff operator to have $\displaystyle\lim_{k\to\infty} \Gamma_k[\phi]=S[\phi] $, 
because of the dimensionful mass being independent of $k$.

\section{Conclusions}
We have proposed an alternative RG flow equation for the effective average action
based on the regularization of the functional integral in space space, 
wherein we perform a balanced coarse-graining procedure by means of the
introduction of a scale dependence in the symplectic form.
Under this regularization both the action and the functional
measure become dependent on a smooth cutoff.
Such a non trivial measure implies the presence of a subtraction term
in the flow equation, as given in
eqns.~\eqref{bosonERGE} and \eqref{GrassmannERGE} for boson and fermion
d.o.f. in quantum mechanics. 
The corresponding RG flow equations in QFT, for a Lorentz 
invariant regularization, are given in eqns.~\eqref{qftbosonERGE} 
and \eqref{qftfermionERGE}. 
The subtraction between the two
traces on the r.h.s. of these equations, induced by the non trivial measure,
gives rise to a better convergence; that is, the r.h.s. could be finite
even if the convergence of the single integrals is not provided by the cutoff operator. 

In flat space, as long as one is not interested in the vacuum energy which is
not observable, and for cutoffs that do not involve any coupling but the field strength, 
nor background fields, the results for this flow are the same as the ones obtained flowing
the standard Wetterich equation.
If however one adopts cutoff schemes more complicated than
the ones we discussed in this work, we cannot exclude some differences
between the flow generated by the present equation, following from phase space
coarse-graining, and the one based on the standard Wetterich
equation. 
This is indeed what we expect in the case of the so called
non pure cutoff schemes that usually include more couplings than just
the field strength.
For example, for a free massive scalar field one could choose
a cutoff scheme involving also the mass term, such as $R_k(-\Box +m^2)$.
Then the flow equation would have a vanishing r.h.s. because of a complete
cancellation between the two traces. Therefore in our framework such
cutoff schemes should be avoided. 
Let us remark that in applications of the Wetterich equation
to the study of matter fields interacting
with gravity, scheme dependences in the beta function of the cosmological
constant were indeed noticed~\cite{CPR,VZ} also in the standard approach.
We plan to address such issues in forthcoming studies.
Also, if the background field method is used, which is very commonly adopted in gauge theories 
in order to preserve gauge invariance, 
the cutoff operator becomes dependent on the background fields.
Therefore the subtraction term induced by the
non trivial measure also depends on them and this leads to an EAA whose background dependence 
differs from the usual one~\footnote{A flow equation for scalar QED modified
by a subtraction term dependent on the background gauge field was
already used in~\cite{RW} with the motivation of minimizing some quantities~\cite{reuter}.}.

We have also discussed the relation which ties the modified EAA
$\Gamma^\Lambda_k$ to the Wilsonian (i.e. bare) action $S^\Lambda$ in the presence
of an UV cutoff $\Lambda$. 
The non trivial functional measure plays here a fundamental role,
providing the correct normalization for a rising functional delta inside
the path integral, which is realized when the IR cutoff scale of the EAA reaches its UV limit. 
We have analyzed how such a UV limit is defined according to the properties
of the cutoff operator which implements the coarse-graining, and classified
some possible ways in which the EAA can approach the Wilsonian action.
Finally we have addressed the same problem in the case in which no UV cutoff
is present. 

In so doing we have computed the contribution to the vacuum energy density of free
massive theories in arbitrary dimensions.
In particular we have found that, under preservation of Lorentz invariance, the
contributions of quantum fluctuations to the vacuum energy density grow
 only quadratically in the UV cutoff and vanish in the massless case.
Such a scenario, i.e the absence of the quartic divergences, was already
invoked recently~\cite{akhmedov,OS} in a standard perturbative QFT approach 
(where infinite constant contributions from the functional measure were
neglected) by performing ad hoc regularizations and subtractions justified by 
symmetry and reality conditions.
We think that our computation gives for the first time a straightforward and neat
derivation of such a behavior which leads to a less dramatic fine tuning problem for the cosmological constant
associated to free quantum fields.
Also, a paradoxical effect about the contribution of the low energy modes
 to the cosmological constant was observed in~\cite{RM} by using the standard Wetterich equation.
With the present equation such an effect disappears; e.g. for bosonic fields such a contribution
grows monotonically for decreasing $k$. 
The method of following the RG flow of the vacuum energy
density, or of the cosmological constant term in a curved spacetime, 
is also suitable for studying the case of interacting theories and
we hope to report on this interesting problem soon.

One of the possible areas of application of this modified flow is the study of
gravity plus matter systems. It will be interesting to trace the possible
differences brought by this approach in the flow and fixed point structures of such
interacting theories. In the absence of an UV cutoff in the theory, which is
the normal setup for such models, the effect of the choice of cutoff operators
with the right singularity structure to provide a convergence to the bare action,
 as discussed in the section $4.2$, is also an important point to be investigated.
Simpler questions related to QFT on curved background spacetimes can
also be addressed in this framework.

Finally it could be very interesting to generalize this regularization prescription 
to the case of nonlinear sigma models because they are already characterized
by a non trivial measure in configuration space, hence the introduction of a
cutoff operator in such a measure is expected to affect several observables.

\vspace{0.5cm}
\noindent {\bf Acknowledgments}

The authors would like to thank O. Zanusso, M. Reuter, R. Percacci  and  F. Bastianelli for
stimulating discussions. 


\end{document}